\documentclass[a4paper, 11pt]{article}

\usepackage{amsfonts,amssymb}
\usepackage{amsmath}
\usepackage{enumerate}
\usepackage{amsthm}
\usepackage{color}
\usepackage{url}
\usepackage[title]{appendix}
\usepackage[utf8]{inputenc}
\usepackage{geometry}
\usepackage{graphicx}
\usepackage{color}
\usepackage{transparent}
\usepackage{setspace}
\usepackage{hyperref}
\usepackage{amsbsy}
\usepackage{float}
\usepackage{authblk}

\author[1]{Martin Bonte\footnote{Corresponding author, e-mail: \href{mailto:b.blah@gmail.com}{martin.bonte@meteo.be}}}
\author[1, 2]{Lesley De Cruz}
\author[1]{Fabian Debal}
\author[3]{Stéphane Vannitsem}
\affil[1]{Royal Meteorological Institute of Belgium, Brussels, Belgium}
\affil[2]{Vrije Universiteit Brussel, Brussels, Belgium}
\affil[3]{School of Physical and Mathematical Sciences \& The Asian School of the Environment, Nanyang Technological University, Singapore}

\title{Spread/Error relationship and spatial error structure of precipitation ensemble nowcasting: Comparison of STEPS and generative AI}

\date{}
\begin{document}

\maketitle

\section*{Abstract}

The predictability of the generative AI-based nowcasting model LDCast (trained on another region) is evaluated over Belgium, together with the pysteps implementation of the nowcasting algorithm STEPS. STEPS and LDCast are slightly underdispersive, but the ensemble spread provides an estimation of the error at almost all scales. Both models adapt the properties of their ensembles to the type of event, either convective or stratiform. The spatial scores of the STEPS and LDCast ensembles are compared with those of surrogate ensembles having some key properties, revealing that both STEPS and LDCast have very little ability to spatially localise the ensemble mean error vector through their ensemble members. This suggests that the content of STEPS and LDCast ensembles is informative in terms of statistics, but not in terms of dynamics.

\section{Introduction}

Many artificial intelligence (AI) models for weather and climate have recently been developed. However, detailed evaluations of these models are still limited and are often done from a statistical point of view, using general scores such as the RMSE between a single forecast (or the ensemble mean) and the observations. In order to be used by forecasters in operational activities, the consistency of the meteorological characteristics of their nowcasts must be assessed \cite{Benbouallegue2024,Radford2025}.

Since the dawn of General Circulation Models (GCMs), there have been many in-depth analyses of their dynamics and predictability in the past 70 years (e.g. \cite{Kalnay2003, Bauer2015}). Recently, Artificial Intelligence Weather Prediction (AIWP) models such as Pangu-Weather, GraphCast, and FourCastNet were developed, and their forecasts for various events of interest have been compared in detail with those from Numerical Weather Prediction (NWP) models \cite{Olivetti2024, CharltonPerez2024, Pasche2025, Hua2025}. \cite{Bano-Medina2025a} also conducted a sensitivity analysis of initial conditions on the Spherical Fourier Neural Operator (SFNO) in a storm environment, finding properties similar to a NWP model.

Methods for estimating uncertainties in AI model forecasts are still under development. Bred vectors for AIWP models were constructed by \cite{Bano-Medina2025b, Mahesh2025, Almeida2025}, and a similar method was developed by \cite{Pu2025}. On the other hand, the uncertainty could also be represented naturally by generative models or shaping Gaussian noise as illustrated in \cite{Lang2024}. These models produce nondeterministic forecasts, which makes it easy to create ensembles of forecasts. Generative models such as DGMR \cite{Ravuri2021} and LDCast can also be trained to produce realistic forecasts that retain sharpness instead of being smoothed out due to the loss of predictability with increasing lead time.

However, the relevance of ensemble properties of generative models for operational activities has not been sufficiently investigated. Indeed, these model ensembles are primarily evaluated over the whole dataset using global scores \cite{Lang2024} or global rank histograms \cite{Price2023, Leinonen2023}. Some models are also evaluated on a few selected events through a meteorologist's comparison and ranking of the outputs of different models \cite{Ravuri2021, Zhang2023}.

These evaluation methods are necessary first steps, but a deeper understanding of the characteristics of generative models is needed. This includes their dynamical properties, as well as their particular behavior and biases for different weather types or spatial scales. The central question of this paper is therefore to characterize the information contained in the ensembles generated by a generative nowcasting model, with a focus on its predictability properties.

This work aims to provide this characterization for LDCast \cite{Leinonen2023}, a generative model for rainfall nowcasting. It is evaluated using the Belgian radar composite RADCLIM \cite{Goudenhoofdt2016, Journee2023}, but it has not been retrained on this dataset. Therefore, the model weights are the original ones, and the model can be considered pre-trained for the Belgian radar precipitation dataset. The pysteps implementation of the nowcasting algorithm STEPS \cite{Bowler2006, Seed2013, Pulkkinen2019} is evaluated in parallel to clarify which approach is best in producing rainfall forecasts for both convective and stratriform rainfall cases.

The three main results for LDCast are the following:
\begin{enumerate}
    \item The spread of LDCast ensembles saturates from small to large scales and provides an estimation of the current error at nearly all scales (Sec. \ref{spectral_error_variance_sec}).
    \item Depending on the event type, LDCast can adapt the power spectra of the perturbation modes of the ensembles, as well as the distribution of the perturbation sizes (Sec. \ref{morphology_sec}).
    \item Ensembles of this version of LDCast are not better at spatially capturing the error than surrogate ensembles with the appropriate statistical properties, for the metrics used in this work (Sec. \ref{spatial_skill_sec}).
\end{enumerate}
These results are also valid for STEPS, with the difference that STEPS ensembles seem to collapse, meaning that the members tend to align along a few directions in phase space.

\section{Results}

In order to analyze the behaviors of STEPS and LDCast in dynamically different situations, ten convective and ten stratiform events were selected (see Sec. \ref{events_selection_preprocessing}) and the ensembles of nowcasts produced by STEPS and LDCast were analyzed. The results are presented below.

\subsection{Spectral error and spectral variance}

\label{spectral_error_variance_sec}

\begin{figure}
\centering
\hspace*{-1cm}
\includegraphics[scale=0.53]{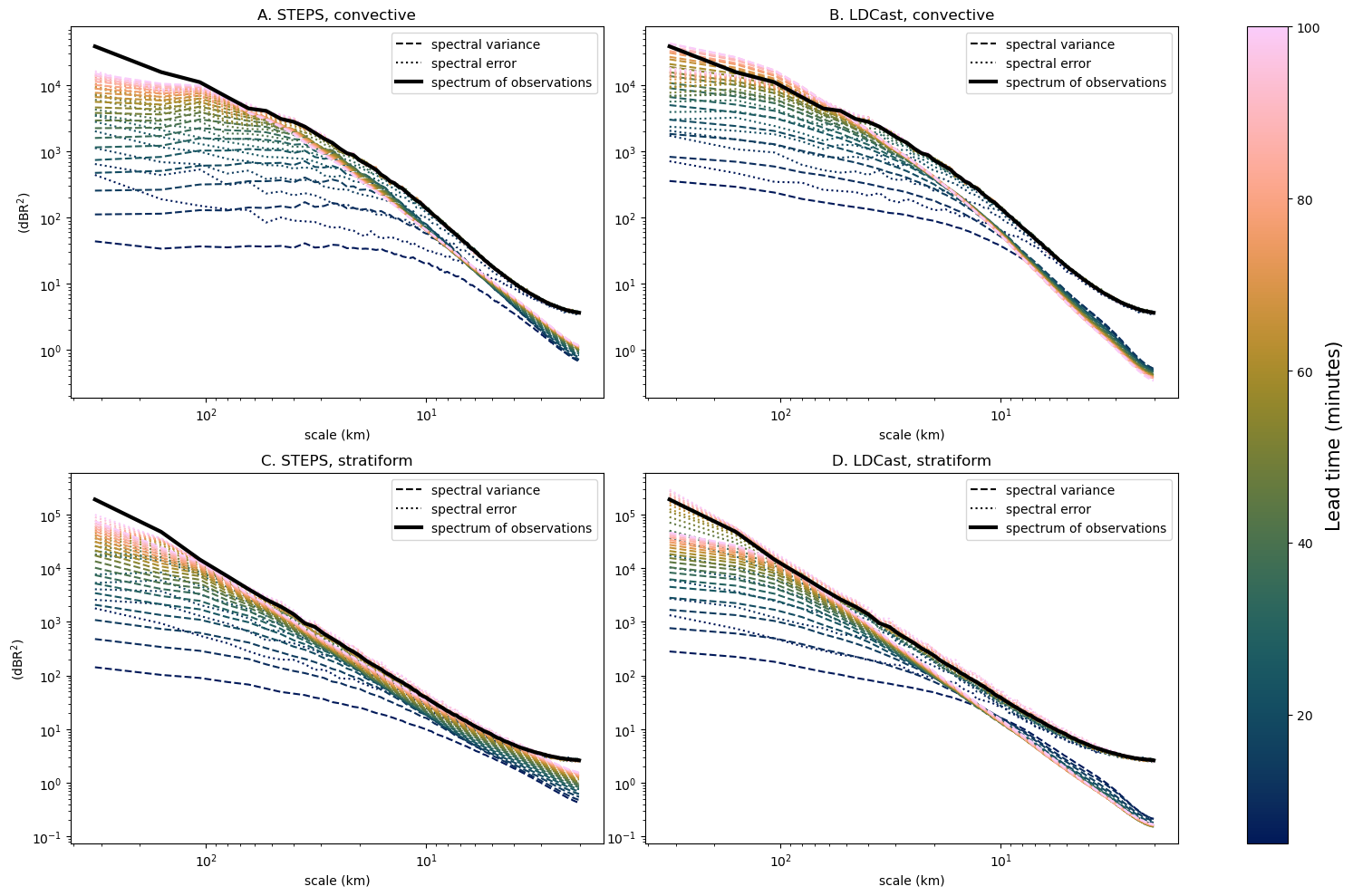}
\caption{Spectral error and spectral variance, for STEPS and for LDCast, and for convective and for stratiform events (averaged over events). Dashed lines display the spectral variance and dotted lines display the spectral error. The color represents the lead time. The power spectra of observations are represented by thick black lines.}
\label{spectral_error_variance_fig}
\end{figure}

The spectral error and spectral variance for STEPS and for LDCast, together with the power spectra of observations, are displayed in Fig. \ref{spectral_error_variance_fig} (average over the selected events). The color of the curves represents the lead time at which the spread and the error are computed.

Both the spectral error and the spectral variance increase with lead time and eventually saturate to the power spectrum of the observation (see Appendix \ref{error_saturation_sec}). The error at some scales is immediately saturated \cite{Pulkkinen2019}. In convective events, those are the scales below 5-6 km for both STEPS and LDCast. Interestingly, there are more scales (up to 9-10km) that are saturated at the beginning of the nowcasts of stratiform events. This is in agreement with the fact that the ensembles of both models have larger-scale perturbations in stratiform events than in convective events (Sec. \ref{morphology_sec}).

\begin{figure}
\centering
\hspace*{-0cm}
\includegraphics[scale=0.5]{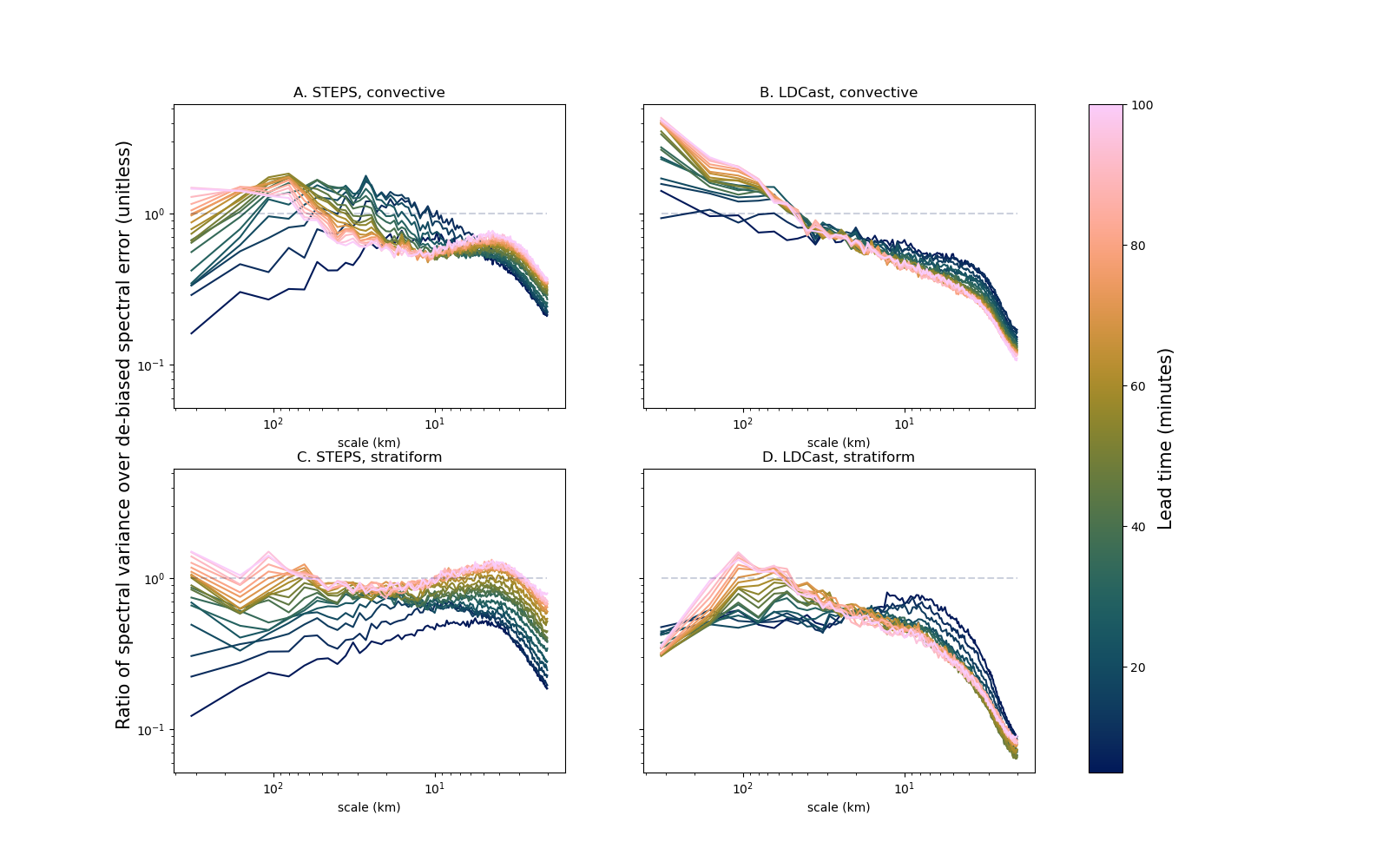}
\caption{Ratios of the spectral variance over the de-biased spectral error, for STEPS and LDCast, and for convective and stratiform events (averaged over the events). The color represents the lead time. Horizontal dashed grey lines mark where the ratio is $1$. Ratio $<1$: underdispersion, ratio $>1$: overdispersion.}
\label{spread-skill_STEPS_LDCast}
\end{figure}

For a well-calibrated ensemble, the error and the spread should be equal, provided the ensemble mean is not biased \cite{Fortin2014}. Whether STEPS and LDCast ensembles are well-calibrated is assessed more precisely with Fig. \ref{spread-skill_STEPS_LDCast}, where the scale-by-scale ratio of the spectral variance over the de-biased spectral error is displayed. A ratio smaller than 1 indicates underdispersion, while there is overdispersion when it is larger than 1.

Overall, both models have variance/error ratios close to 1 for scales above 5 km. LDCast is slightly underdispersive, except for the larger scales in convective events, where it is rather overdispersive. On the other hand, for stratiform events, STEPS ensembles are underdispersive for short lead times but become well calibrated for longer lead times. In convective cases, STEPS is either underdispersive or overdispersive depending on the lead time and the scale.

For scales below 5 km, all ensembles seem to be underdispersive. However, the power spectrum of radar images is known to flatten due to white noise contamination at these scales due to non-meteorological sources \cite{Seed2013}, and that explains the shape of the corresponding part of the spectral error. Therefore, the flattening of the power spectrum of radar images can be considered as not physical, and this underdispersion should not be identified as a weakness of the ensembles. This occurs at scales for which the error is immediately saturated, so that this is anyway irrelevant from a forecasting point of view.

\subsection{Morphology of STEPS and LDCast ensembles}

\label{morphology_sec}

\begin{figure}[H]
\centering
\includegraphics[scale=0.5]{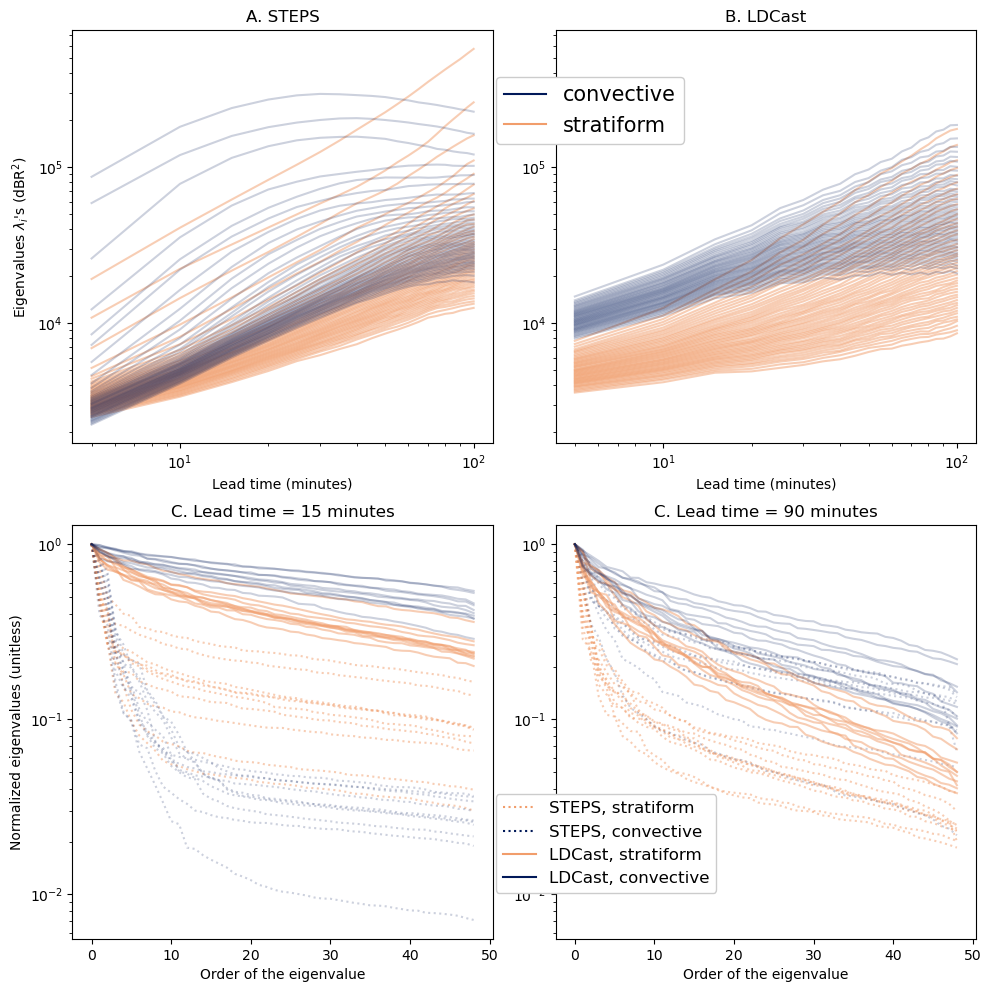}
\caption{Eigenvalues $\lambda_i$ of the covariance matrix. Evolution of the eigenvalues for convective and stratiform events for STEPS (A.) and LDCast (B.). Normalized eigenvalues for convective and stratiform events, for STEPS and LDCast, after 15 minutes (C.) and after 90 minutes (D.) (each line is the eigenvalue spectrum for one event).}
\label{error_growth_fig}
\end{figure}

Panels A. and B. of Fig. \ref{error_growth_fig} display the eigenvalues $\lambda_i$ of the covariance matrix for both models, averaged over events (see Sec. \ref{covariance_seq}). A few eigenvalues largely dominate the others, especially in convective STEPS ensembles (panel A.). To investigate the geometry of these ensembles, the residual vectors $v_i$ of the members with respect to the ensemble mean were considered (see Eq. \eqref{residual_vec_eq}). There are roughly speaking two different geometries that could explain this: a) the $v_i$ have comparable norms but they align along the directions of the first eigenvectors of the covariance matrix or b) the $v_i$ are each pointing in different directions but some of them have much larger norms. In order to determine which of the two geometries is actually realized in this case, the cosine $c_{ij}$ of the angle between the residual vector $v_i$ and the $j$th eigenvector was computed for all $i$ and $j$. 

\begin{figure}[H]
\centering
\includegraphics[scale=0.5]{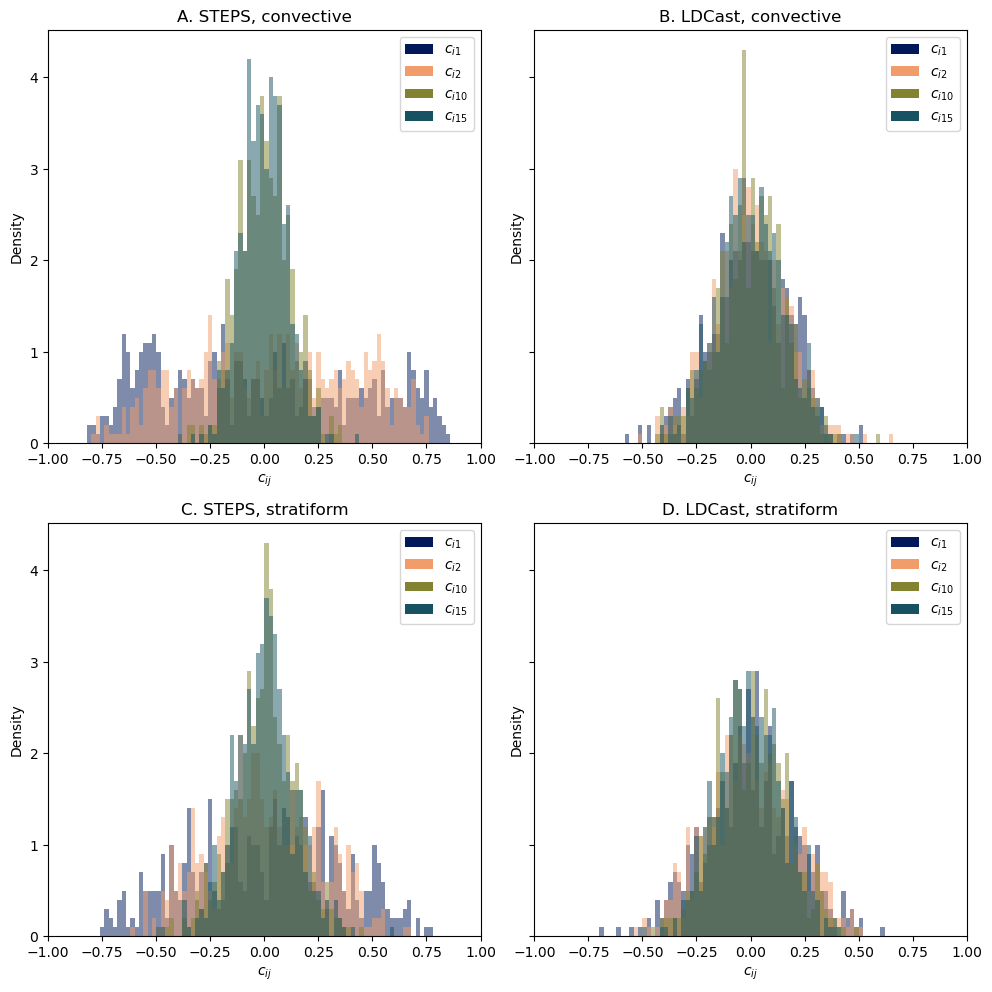}
\caption{Histograms of the $c_{ij}$ for $j = 1, 5, 10$ and $15$ (for all events) after 15 minutes.}
\label{hist_projections_fig}
\end{figure}

Figure \ref{hist_projections_fig} represents the histograms of values of $c_{ij}$ for $j = 1, 2, 10$ and $15$, for a lead time of 15 minutes. In convective STEPS ensembles (panel A.), the distributions of $c_{i1}$ and $c_{i2}$ contain much higher values than the distributions of $c_{i10}$ and $c_{i15}$, meaning that the $v_i$ project much more on the first and the second eigenvectors than on the 10th and the 15th. This is also the case for stratiform STEPS ensemble, albeit to a lesser extent. The conclusion is that the $v_i$ in these ensembles align along the first eigenvectors (option a)).

The evolution of eigenvalues $\lambda_i$ of the covariance matrix for stratiform events is comparable for STEPS and LDCast: the eigenvalues increase steadily following power laws of the lead time (Fig. 3, panels A. and B.). This is also the case for convective LDCast ensembles. These power laws should be connected with the theoretical time that an error, initially located at small scales, needs to reach a given larger scale. Indeed, on a theoretical basis, it is expected that the highest scale contaminated by the error evolves as an exponential of the lead time 2D turbulent fluids (as is seen when working on synoptic scales), while the highest scale reached by the error increases as a power law of the lead time for 3D turbulent fluids \cite{Vallis2017}. Given the scales of nowcasting, the relevant physical processes are those of 3D turbulence, so that there is a qualitative agreement between these theoretical considerations and the scaling on the error observed in Fig. \ref{error_growth_fig}.

Panels C. and D. of Fig. 3 show, for two different lead times, the spectrum of the normalized eigenvalues (which are the eigenvalues divided by the highest one in the ensemble). Consistently with the panels A. and B., it appears that LDCast ensembles have generally more homogeneous eigenvalues spectra (i.e. have more eigenvalues of similar size) than STEPS ensembles. Convective LDCast ensembles exhibit more homogeneous eigenvalues spectra than stratiform LDCast ensembles, especially at long lead times. As already noted from panel A., it is also obvious that STEPS ensembles are largely dominated by a few eigenvalues: convective STEPS ensembles are dominated  by $\sim 10$ eigenvalues, while stratiform STEPS ensembles are dominated by $\sim 5$ eigenvalues. However, the relative size of higher order eigenvalues (order $> 5$ in stratiform ensembles, order $> 10$ in convective ensembles) is larger in stratiform STEPS ensembles than in convective STEPS ensembles.

\begin{figure}
\centering
\hspace*{-0.5cm}
\includegraphics[scale=0.5]{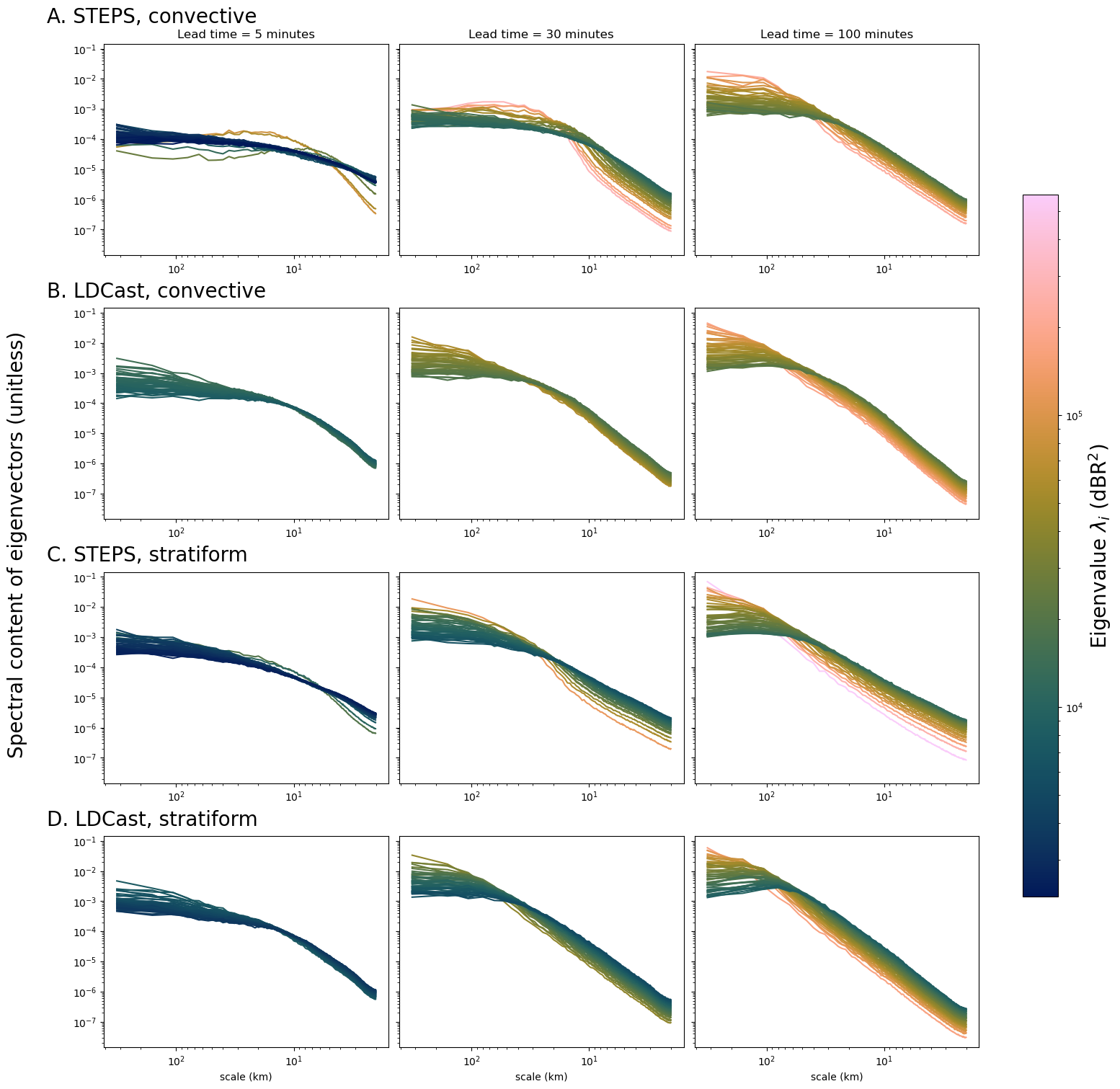}
\caption{Evolution of the spectra of the eigenvectors for convective and stratiform events, for LDCast and STEPS ensembles. The spectra are averaged over all events. The color represents the corresponding eigenvalue.}
\label{uncertainty_modes_fig}
\end{figure}

The power spectra of the eigenvectors of the covariance matrix are displayed in Fig. \ref{uncertainty_modes_fig}, for different lead times, for stratiform and convective events. The color represents the associated eigenvalues $\lambda_i$. Overall, the perturbation modes gradually develop over larger scales, while the spectrum profile remains the same for small scales. This shows that the directions of perturbation evolve in phase space with the lead time: at long lead times, the perturbations are vectors pointing in directions corresponding to larger scales than at the beginning of the nowcast.

It also appears that eigenvectors with higher eigenvalues have more power at large scales. This leads to a hierarchy: perturbations with larger amplitudes affect more the large scales. This is in agreement with the fact that, for a given scale, there is a maximum size for the error (see Fig. \ref{spectral_error_variance_fig}), so that once a perturbation reached the saturation value at a given scale, it propagates to larger scales.

For both STEPS and LDCast, the eigenvectors of stratiform ensembles (especially the eigenvectors associated with the highest eigenvalues) have more power at larger scales than in convective ensembles. This means that the perturbations contained in stratiform ensembles are at larger scales than those in convective ensembles.

The curves standing out in the top row (A.) of Fig. \ref{uncertainty_modes_fig} for short and intermediate lead times are those of the power spectra of the eigenvectors that largely dominate the STEPS ensembles in convective cases (panel A. of Fig. \ref{error_growth_fig}). Similar power spectra are also present in stratiform STEPS ensembles.

Spectra at small scales decrease over time. This is because the eigenvectors have unit norm, so that the sum over the scales of each power spectrum is equal to $1$.

\subsection{Spatial skill of STEPS and LDCast ensembles}

\label{spatial_skill_sec}

This section presents results assessing the skill of ensemble members to spatially represent the error. The focus is particularly on evaluating whether the ensembles contain any spatial information beyond that present in the ensemble mean. 

Two metrics are considered. The first is $\cos\gamma(e, p)$ and is the cosine of the angle between the error $e$ and its projection $p$ on the subspace spanned by the residual vectors $v_i$ (a similar quantity was considered in \cite{Uboldi2015}), and the second is the probabilistic Fraction Skill Score (FSS) \cite{Schwartz2010}. See Sec. \ref{spatial_metrics_sec} for more details on these two metrics. The two metrics are computed for STEPS and LDCast ensembles and compared to those of MAAFT ensembles, a baseline of surrogate nowcasts with the same distribution of rainfall intensities and whose residual vectors have the same power spectra.

\subsubsection*{Using $\cos\gamma(e, p)$}

\begin{figure}
\centering
\hspace*{-1.9cm}
\includegraphics[scale=0.6]{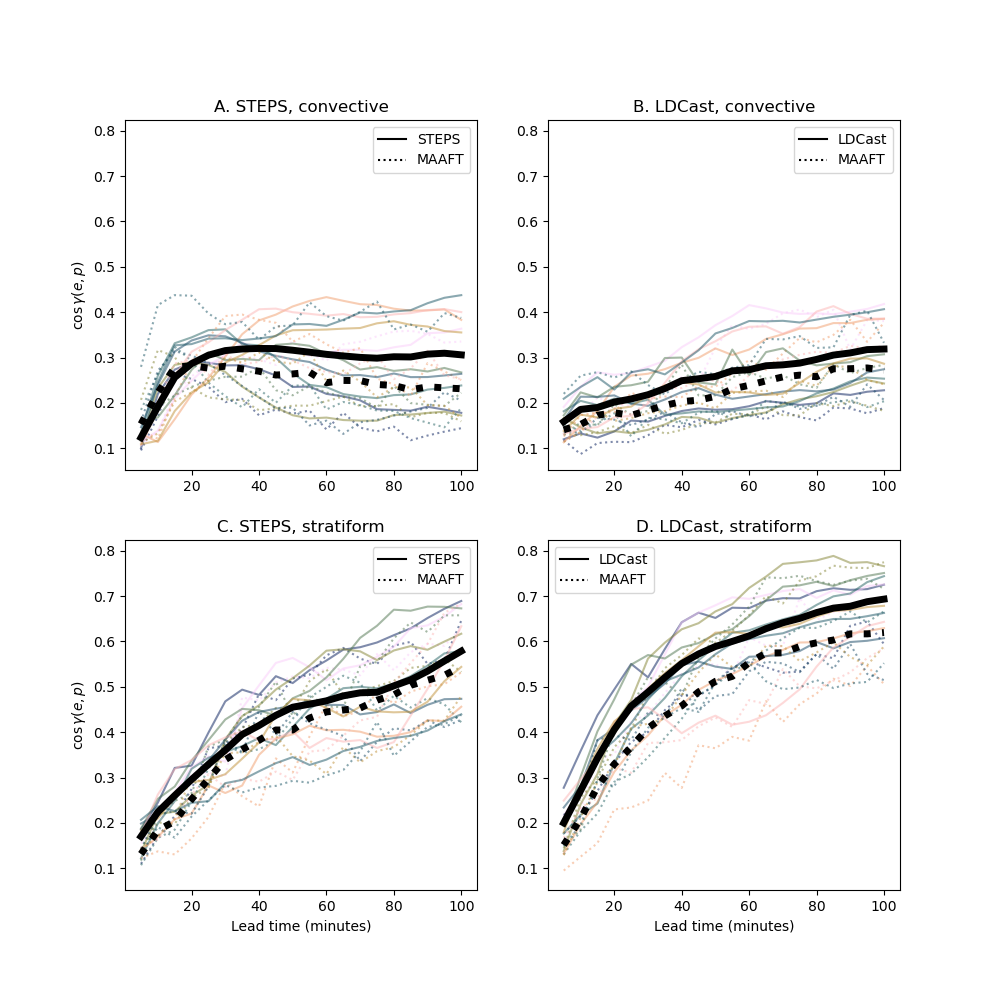}
\caption{Comparison of $\cos\gamma(e, p)$ of the error and its projection on STEPS/LDCast ensembles (solid lines) with the $\cos\gamma(e, p)$ for the corresponding MAAFT ensembles (dotted lines), for all lead times. Thin lines depict $\cos\gamma(e, p)$ for each event and thick lines represent the event average. The MAAFT ensembles whose scores are presented in panels A. and C. are built from the corresponding STEPS ensembles, while those whose scores in presented in panels B. and D. are built from the corresponding LDCast ensembles.}
\label{cosine_fig}
\end{figure}

 The cosine of the angle between the error $e$ and its projection $p$ on the STEPS and LDCast residual vectors $v_i$ is represented by solid lines in the panels of Fig. \ref{cosine_fig}. The thin lines depict this cosine for each event, and the thick line shows the mean of this quantity over the events. The dotted lines represent the same quantity for the MAAFT ensembles constructed from the original STEPS and LDCast ensembles.

The $\cos\gamma(e, p)$ values of the surrogate MAAFT ensembles are very close to those of the corresponding STEPS and LDCast ensembles. The reason for this is the shape of the Fourier spectra of the perturbations. Indeed, it was verified that SPEC ensembles (surrogates generated without the constraint on the distribution of values, only imposing the power spectra of residual vectors, see Sec. \ref{surrogate_sec}) lead to similar values for $\cos\gamma(e, p)$ (Fig. S5 of Supplementary material). In particular, the spatial localization of the perturbations contained in the $v_i$ of STEPS and LDCast ensembles does not explain the values of $\cos\gamma(e, p)$ of the error with these ensembles, since the same values are reached for surrogate ensembles with random residual vectors where no constraint is imposed on the spatial localization.

One way to understand this is by expressing $\cos\gamma(e, p)$ in terms of Fourier modes: if $E_{\vec{k}}$ and $V_{\vec{k}}$ are respectively the Fourier coefficients of $e$ and of a residual vector $v$ (of a STEPS, LDCast or MAAFT ensemble) and $\phi_{\vec{k}}^e$ and $\phi_{\vec{k}}^v$ the complex phases of these coefficients, the terms in the projection of $e$ on the $v_i$ is proportional to the scalar product of $e$ and $v$, which is of the form $|E_{\vec{k}}||V_{\vec{k}}|\left(\phi_{\vec{k}}^e\phi_{\vec{k}}^{v*} + \phi_{\vec{k}}^{e*}\phi_{\vec{k}}^{v}\right)$. The moduli $|V_{\vec k}|$ are imposed when constructing MAAFT ensembles and the complex phases $\phi_{\vec k}^v$ are precisely random. The fact that the values of $\cos\gamma(e, p)$ for a STEPS or an LDCast ensemble are the same as with the corresponding MAAFT ensemble shows that the complex phases $\phi_{\vec k}^v$ are also random for STEPS and LDCast ensembles. Since the spatial localization of the structures of a field is contained in the complex phases of its Fourier coefficients, the fact that the complex phases for the Fourier coefficients of the $v_i$ are random indeed means that the localization of the structures they contain is random.

The value of $\cos\gamma(e, p)$ increases with the lead time and this can be explained by the evolution of the power spectra of the vectors $v_i$. When the lead time increases, these vectors progressively represent larger perturbations. Because of this, it can be expected that it becomes easier for ensembles to cover the whole domain with independent perturbations of a given scale, and thus to spatially capture the actual instability.

Interestingly, \cite{Feng2024} argued that the error at convective scales always has a negligible projection onto the perturbation modes of an ensemble. While this is always the case at short lead times in Fig. \ref{cosine_fig}, the projection is largely non-negligible for intermediate and long lead times, especially for stratiform events. However, this is because the perturbation modes develop large-scale components that eventually cover the entire domain.

\subsubsection*{Using the FSS}

\begin{figure}
\centering
\hspace*{-0cm}
\includegraphics[scale=0.47]{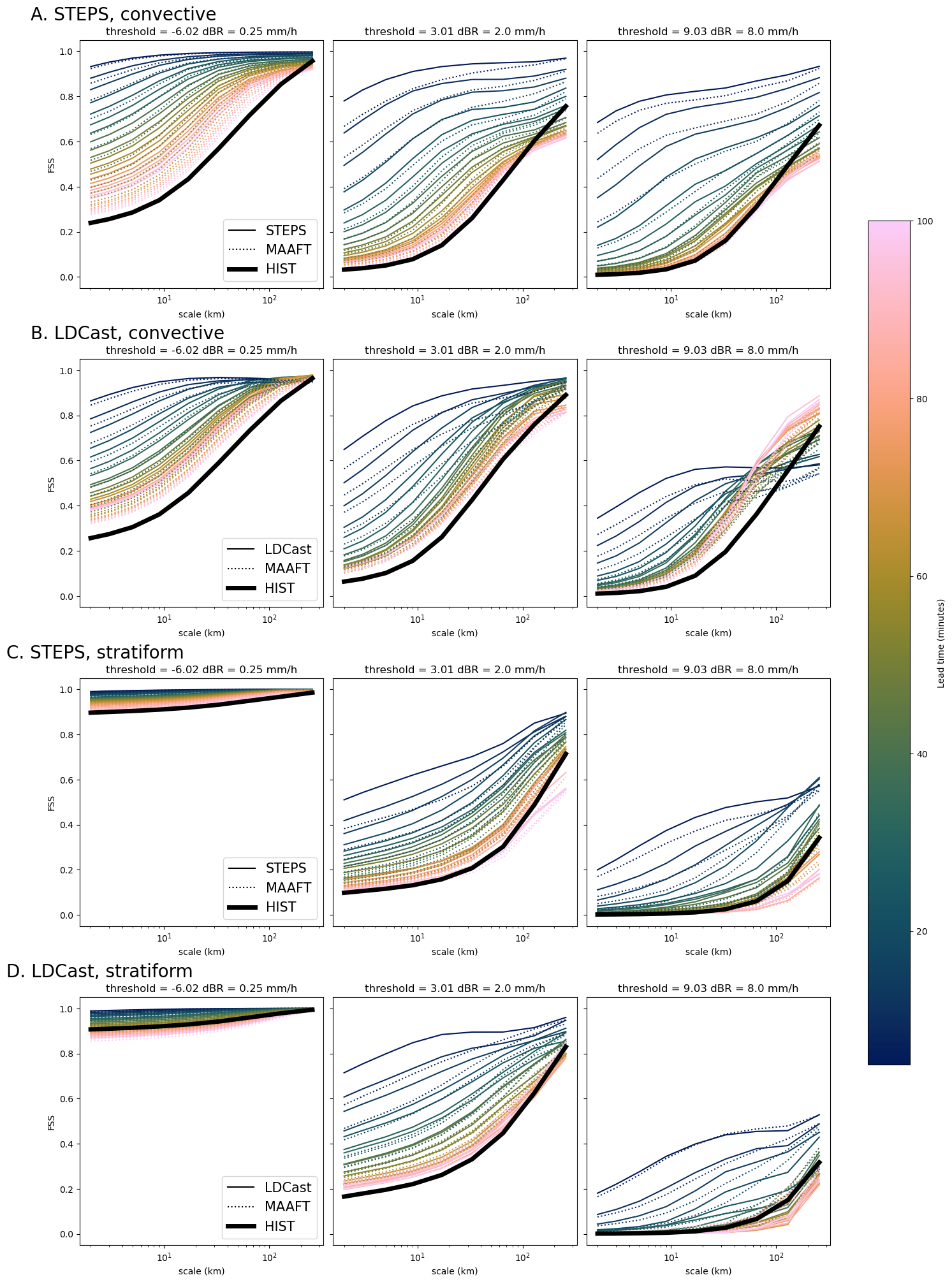}
\caption{Fraction Skill Scores (FSSs) as functions of scale for a) convective events and b) stratiform events, with the time dependence represented by the color. The thin solid lines represent the LDCast FSSs, the dotted lines represent the MAAFT FSSs while the thick solid line represents the HIST FSSs (no time dependence shown). The $x$-axis represents the scale for which the FSS was computed, and is reversed with respect to the previous Figures.}
\label{FSS_fig}
\end{figure}

Figure \ref{FSS_fig} depicts the FSS averaged over events, for STEPS/LDCast ensembles and for the corresponding MAAFT ensembles, by solid and dotted lines, respectively. The thick black lines (HIST lines) in this Figure represent the FSS of the surrogate HIST ensemble (surrogates constructed by simply shuffling the pixel values, without power-spectrum adjustment, see Sec. \ref{surrogate_sec}). This score remains relatively constant over time, so that only the time average is represented in Fig. \ref{FSS_fig}. This FSS is the one of a random prediction given the distribution of values in the nowcast. For the smallest scales, this FSS is analoguous to the random FSS in \cite{Roberts2008}. For the largest scale, it corresponds to their Asymptotic Fraction Skill Score (AFSS) \cite{Roberts2008}.

The LDCast and the corresponding MAAFT scores saturate with lead time to a value close to the HIST value. Interestingly, the LDCast and STEPS values of the FSS saturate more slowly than the spectral variance in Fig. \ref{spectral_error_variance_fig}: even at the smallest scales, the FSS saturates after 40 minutes, whereas the spectral variance is immediately saturated at those scales. The other intriguing feature in this Figure is the behavior of LDCast for the highest threshold: for the highest scales for which it is computed, the FSS scores increase with lead time. At the highest scale, the FSS score merely assesses whether there is a correct number (over the whole radar frame) of values above the chosen threshold. In this case, the number of high-intensity values is better estimated as the lead time increases.

The MAAFT FSSs are quite close to STEPS and LDCast FSSs, showing again that the FSS scores of STEPS and LDCast ensembles are mainly due to their statistical properties, but not to the spatial localization of the perturbations in the ensembles.

\section{Discussion}

\label{summary}

This work investigates spectral and spatial properties of the ensembles of STEPS and of a pre-trained version of LDCast. The spreads of ensembles of both models saturate from small to large scales and overall, they provide an estimation of the error for most scales. The latter observation extends the conclusion of \cite{Leinonen2023} in the case of LDCast. Depending on the event type, STEPS and LDCast adapt the perturbation modes of their ensembles, as well as the distribution of perturbation sizes. STEPS ensemble members tend however to align along a few directions in phase space. This implies that STEPS ensembles have only a few modes of variability, and that the members are quite similar to each other.

Spatial scores were computed to assess STEPS's and LDCast's ability to capture the error in the ensembles. The surrogate MAAFT ensembles were designed to have similar statistical properties to those of original ensembles, allowing to challenge the scores of the models. MAAFT ensembles and original ensembles have very close scores, demonstrating that the skill of STEPS and this version of LDCast is largely due to the statistical properties of the ensembles, and not to the localization of the error and spread. Appropriate power spectra are sufficient to reproduce the values of the cosine of the angle between the error and original ensembles, while an adequate distribution of pixel values is also needed to reproduce the FSS.

According to the metrics used in this work ($\cos\gamma(e, p)$ and the FSS), these results show that the ensembles of STEPS and LDCast, at least without retraining the latter, do not have any dynamical information on the spatial localization of the error.

The domain on which the models were evaluated was chosen to be smaller than the region over which the nowcasts were computed (see Sec. \ref{models}) in order to avoid having missing values in STEPS nowcasts. However, the fact that STEPS has missing values in its nowcasts can also be seen as an advantage of the model: the missing values show explicitly where information was missing to produce the nowcast. In contrast, it is more difficult to spot where a generative model like LDCast has completely invented the values.

There is a number of possible directions open for future work. First, it should include the same type of analysis for a version of LDCast that has been trained over the Belgian region. Other generative models for nowcasting should also be evaluated. It would be interesting to see if the ensembles also have similar properties in Fourier space. This might not be the case for all models, depending on their architectures. The version of pysteps blended with NWP forecasts \cite{Imhoff2023} would be also interesting to evaluate.

The particular structure of STEPS ensembles should also be further investigated, to understand the origin of the alignment of the $v_i$ and whether some correction should be applied. This work focused on the spatial localization of the perturbations, but did not evaluate the spatial correlations in the field themselves. It is important for some downstream applications (such as hydrological simulations) to have nowcasts correctly representing the current anisotropy of the rainfall fields in the ensemble members. The nonparametric method for generating noise \cite{Seed2013} in STEPS ensembles allows to generate members with the appropriate spatial correlation, but it is not known how LDCast members are in this regard. The results in this study point to the fact that, in the case where the spatial anisotropy is correctly represented in the nowcasts (which  is the case in STEPS nowcasts when the nonparametric method is used), the structures are not better localized for the members than for the ensemble mean.

The mechanism proposed in Sec. \ref{spatial_skill_sec} to explain the increase of $\cos\gamma(e, p)$ over time (Fig. \ref{cosine_fig}) seems to imply that there is a relationship between the ensemble size, the domain size and the scale at which $\cos\gamma(e, p)$ saturates. It would be interesting to quantify this relationship more precisely by varying the ensemble size.

\section{Methods}

\label{models_methods}

\subsection{Data and preprocessing}

\label{events_selection_preprocessing}

The radar product on which STEPS and LDCast are evaluated in this work is RADCLIM \cite{Goudenhoofdt2016, Journee2023}: it is a quantitative precipitation estimation product based on radar measurements, which are merged with rain gauge measures. Several techniques exist for this merging: the product used in this study was obtained with the Kriging with External Drift (KED) technique. The time resolution is 5 minutes and the spatial resolution is 1 km $\times$ 1 km. The evaluation domain is a 320 px$\times$ 320 px square in the radar frame.

Ten stratiform and ten convective events were selected based on the convective precipitation and the mean large-scale precipitation derived from ERA5 \cite{Hersbach2023}. Events are considered convective if their mean convective precipitation was greater than $10^{-4}$ mm/h and their large-scale precipitation was less than $10^{-6}$ mm/h. Similarly, stratiform events were selected as those with a mean convective precipitation below $10^{-7}$ mm/h and a large-scale precipitation above $8.07 \cdot 10^{-4}$ mm/h.

As both STEPS and LDCast transform the rain rate to a logarithmic scale, all computations in this work are done with a similar scale. The transformation used in this work is
\begin{align}
f(R) = \left\{ \begin{array}{cc}
10\,\mathrm{log}_{10}(R) & \text{for } R \geq 0.1 \text{ mm/h}\\
- 15 & \text{for } R < 0.1 \text{mm/h}
\end{array}\right.
\end{align}
The unit after this transformation is dBR.

\subsection{Models}

\label{models}

The two nowcasting models considered in this work are STEPS and LDCast. STEPS is built on Lagrangian persistence \cite{Zawadzki1994}, which consists in advecting the rainfall field with a motion field estimated through an optical flow algorithm. In STEPS, the rainfall field is in addition decomposed into fields of different spatial scales via a cascade, and each of the levels of the cascade evolves according to an auto-regressive (AR) process, usually of order 2. There are two sources of perturbation in STEPS: the motion field is stochastically perturbed via the BPS method, and the field intensities are perturbed via the noise component of the AR(2) processes \cite{Bowler2006}. The noise itself was originally generated with a parametric method, but the nonparametric method developed in \cite{Seed2013} is now more commonly used. STEPS is implemented at the RMI with pysteps \cite{Pulkkinen2019}, and features blending with NWP forecasts \cite{Imhoff2023}.

In STEPS, there is no value for the pixels for which the value should come from the advection of pixel values out of the radar frame. To avoid as much as possible this problem, STEPS nowcasts were produced using the whole available radar domain of RADCLIM (see Sec. \ref{events_selection_preprocessing}), and the analysis presented in this work was performed on a smaller 320 px $\times$ 320 px square extracted from this domain. There were still some pixels with missing values after doing this, and these were assigned the no-rain value of $-15$ dBR. In stratiform STEPS ensembles, the fraction of those pixels in an ensemble at a given lead time is at most $3\%$ and it is smaller than $1\%$ for convective STEPS ensembles. The effects of such missing values in STEPS nowcasts are therefore supposed to be negligible.

LDCast is a latent diffusion model, meaning that a sequence of rainfall fields is first encoded to a latent space learned by the variational autoencoder of the model. The forecaster network then predicts the latent representation of future rainfall fields. Conditionally to this output, the denoiser stack produces different members, which are finally transformed back from the latent space to rainfall fields \cite{Leinonen2023}.

The LDCast nowcasts were first produced on a 416 px $\times$ 416 px extracted from the radar domain. The nowcasts were then cropped to the square 320 px $\times$ 320 px square used to evaluate the models.

Examples of nowcasts produced with STEPS and LDCast, as well as well a spatial representation of the mean error and of the spatial standard deviation in the ensembles, are provided in the Supplementary material (Figures S1, S2, S3 and S4).

\subsection{Spectral error and spectral variance}

For each event and for each lead time, an ensemble of nowcasts is produced with STEPS, and another one with LDCast. Each ensemble has 50 members: $\{x_i\}_{i = 1, ..., N}$ with $N = 50$. The ensemble mean is computed pixel-wise as
\begin{align}
\bar{x} = \frac{1}{N} \sum_{i=1}^N x_i.
\end{align}
The time dependence is omitted in the notation for simplicity. The observations (radar images) are denoted by $y$, and the pixel-wise error is
\begin{align}
e = y - \bar{x}.
\label{error_eq}
\end{align}
The residual vectors $v_i$ of the members with respect to the mean are defined as
\begin{align}
v_i = x_i - \bar{x}.
\label{residual_vec_eq}
\end{align}
The power spectrum of a field $x$, at a scale $1/k$, is
\begin{align}
    PS(x)_k = \langle |X_{\vec k}|^2\rangle_{|\vec k| = k},
\end{align}
where $X_{\vec k}$ is the Fourier coefficient of $x$ at wavevector $\vec k$ and $\langle \cdot \rangle_{|\vec k| = k}$ denotes the average over the wavevectors with norm equal to $k$.

The spectral error is computed as the power spectrum of the pixel-wise error $e$. It is a scale-by-scale decomposition of the mean squared error of the ensemble mean. On the other hand, the spectral variance is defined as
\begin{align}
    \sigma^2_k = \frac{1}{N-1}\sum_{i=1}^N PS(x_i - \bar{x})_k
    \label{spectral_variance_eq}
\end{align}
and is often called the spread in meteorological applications. The spread/error ratio in Fig. \ref{spread-skill_STEPS_LDCast} was computed with the de-biased error, meaning that a small bias of the ensemble mean was removed in Eq. \eqref{error_eq}.

\subsection{Covariance matrix of ensembles}

\label{covariance_seq}

The morphology of the ensembles is investigated through their covariance matrix. This matrix is largely singular since the ensemble members have $320^2 = 102400$ components, while the ensembles have only $50$ members. Actually, the covariance matrix has only $49$ non-zero eigenvalues because it is constructed out of the residual vectors $v_i$, which satisfy $\sum_i v_i = 0$ (and it was supposed that this is the only relationship between them).

The eigenvalues of the covariance matrix are denoted $\lambda_i$, and the eigenvectors $u_i$ are also called perturbation modes (PM). The normalized eigenvalues are also considered in Sec. \ref{morphology_sec}: these are the eigenvalues divided by the highest eigenvalue in the ensemble for a given lead time. Considering this quantity allows to better understand the relative sizes of the perturbation modes.

Section \ref{morphology_sec} also exploits the information contained in the cosines of the angle between $v_i$ and the $u_j$ of the covariance matrix. It is computed as 
\begin{align}
    c_{ij} = \frac{v_i\cdot u_j}{|v_i|}
\end{align}
since the eigenvector $u_j$ has a unit norm.

\subsection{Spatial metrics}

\label{spatial_metrics_sec}

Two metrics are considered. The first is $\cos\gamma(e, p)$ is the cosine of the angle between the error $e$ and its projection $p$ on the subspace spanned by the residual vectors $v_i$. Using the fact that the eigenvectors of the covariance matrix are orthonormal and span the same subspace as the $v_i$, $p$ can be written as
\begin{align}
    p = \sum_i (u_i \cdot e) u_i
\end{align}
and its norm is $|p| = \sqrt{\sum_i (u_i \cdot e)^2}$. So $\cos\gamma(e, p)$ can be computed in terms of the scalar product between $e$ and $p$ as
\begin{align}
    \cos\gamma(e, p) = \frac{e\cdot p}{|e||p|},
\end{align}
or in terms of the ratio between the norm of $p$ and the norm of $e$:
\begin{align}
    \cos\gamma(e, p) = \frac{|p|}{|e|}
\end{align}
since $e\cdot p = \sum_i (u_i \cdot e)^2 = |p|^2$. These expressions show that $\cos\gamma(e, p)$ indicates the extent to which the error is captured by the ensemble members. A similar quantity was considered in \cite{Uboldi2015} to estimate the ability of the breeding vectors of a realistic model to capture the error growth in a convective situation.

The second metric considered is the Fraction Skill Score (FSS), which was introduced by \cite{Roberts2008} as a metric for the spatial accuracy of a deterministic forecast. With respect to the mean square error, it mitigates displacement errors by comparing the forecast and the observation over neighborhoods of different scales. \cite{Necker2024} recently compared different versions of the FSS for ensemble verification, and recommended the 'probabilistic FSS' proposed in \cite{Schwartz2010}, which is the one used in the current work.

\subsection{Surrogate ensembles}

\label{surrogate_sec}

This study introduces the MAAFT technique to construct surrogate ensembles. It is inspired by the Iterated Amplitude-Adjusted Fourier Transform (IAAFT) \cite{Schreiber1996}, which is a method to construct random surrogate time series with the same distribution of values and the same autocorrelation function as the original time series. The Modified Amplitude-Adjusted Fourier Transform (MAAFT) produces, for each original ensemble, a surrogate ensemble whose members have a) distributions of values prescribed by the corresponding member in the original ensemble and b) residual vectors with respect to the mean with the same power spectra as the $v_i$.

The MAAFT ensembles are constructed member by member as follows. For each ensemble member $x_i$, a surrogate $z_i$ is initialized by shuffling all its pixel values. The surrogate is then iteratively refined by repeating the following steps:
\begin{enumerate}
    \item compute the residual of $z_i$ with respect to the mean: $w_i = z_i - \bar{x}$
    \item adjust the power spectrum of $w_i$ to that of $v_i = x_i - \bar{x}$: the Fourier coefficients $W_{i, \vec{k}}$ of $w_i$ are replaced by $|V_{i, \vec{k}}| \phi_{\vec{k}}$, where $V_{i, \vec {k}}$ are the Fourier coefficients of $v_i$ and $\phi_{\vec{k}} = W_{i, \vec{k}}/|W_{i, \vec{k}}|$ contains the complex phase of $W_{i, \vec {k}}$
    \item construct the new version of the surrogate as $z_i = w_i + \bar{x}$
    \item adjust the distribution of values of $z_i$ to that of $x_i$: the highest value in $z_i$ is replaced by the highest value in $x_i$, the second highest in $z_i$ by the second highest in $x_i$, and so on.
\end{enumerate}
These four steps were repeated 30 times to construct the MAAFT ensembles in this work.

The only difference from the original IAAFT algorithm is that the power spectrum and the distribution adjustments are not done on the same quantity: the distribution is adjusted on the surrogate member $z_i$ itself, while the power spectrum is adjusted on the residual $w_i = z_i - \bar{x}$ of the surrogate with respect to the ensemble mean. This allows to construct surrogate ensembles with the same power spectrum as the $v_i$ and the same distribution of values for each member while keeping the information of the mean in the ensemble.

The power spectrum adjustment technique is already used in nowcasting contexts, for example to create the noise of the AR(2) processes in the STEPS nowcasting algorithm \cite{Seed2013}.

Another type of surrogate ensemble considered in this work is one where only the power spectrum of the residual vectors is preserved (SPEC ensembles). They are constructed by initializing $z_i$ as in MAAFT, and then performing only once steps 1., 2. and 3. of the MAAFT construction.

The last type of surrogate ensembles (HIST ensembles) used in this work are produced simply by shuffling the pixel values member by member. This produces ensemble members with exactly the same histogram of values, but without any spatial information. The MAAFT ensembles can be seen as ensembles combining the properties of both SPEC and HIST ensembles.

\section{Acknowledgements}

This research has been supported by the Belgian Federal Science Policy Office (BELSPO) under contract number B2/233/P2/PRECIP-PREDICT and through the FED-tWIN programme (Prf-2020-017).

\section{Data availability}

The RADCLIM dataset and the nowcasts will be made available upon reasonable requests to the authors.

\section{Code availability}

The code used in this work is available at the Zenodo repository \url{https://doi.org/10.5281/zenodo.18341086}.

\section{Author contributions}

M.B., L.D.C. and S.V. designed the study. M.B. wrote the code and performed the computations. M.B., L.D.C. and S.V. interpreted the results and wrote the manuscript. F.D. helped in the selection of convective and stratiform events.

The corresponding author is Martin Bonte.

\section{Competing interests}

The authors have no competing interests.

\begin{appendices}

\section{Error saturation}

\label{error_saturation_sec}

Both the spectral error and spectral variance in Fig. \ref{spectral_error_variance_fig} reach a maximum scale-dependent value during the nowcast. This saturation value is close to the power spectrum of the rain field itself.

\begin{figure}
\centering
\includegraphics[scale=0.5]{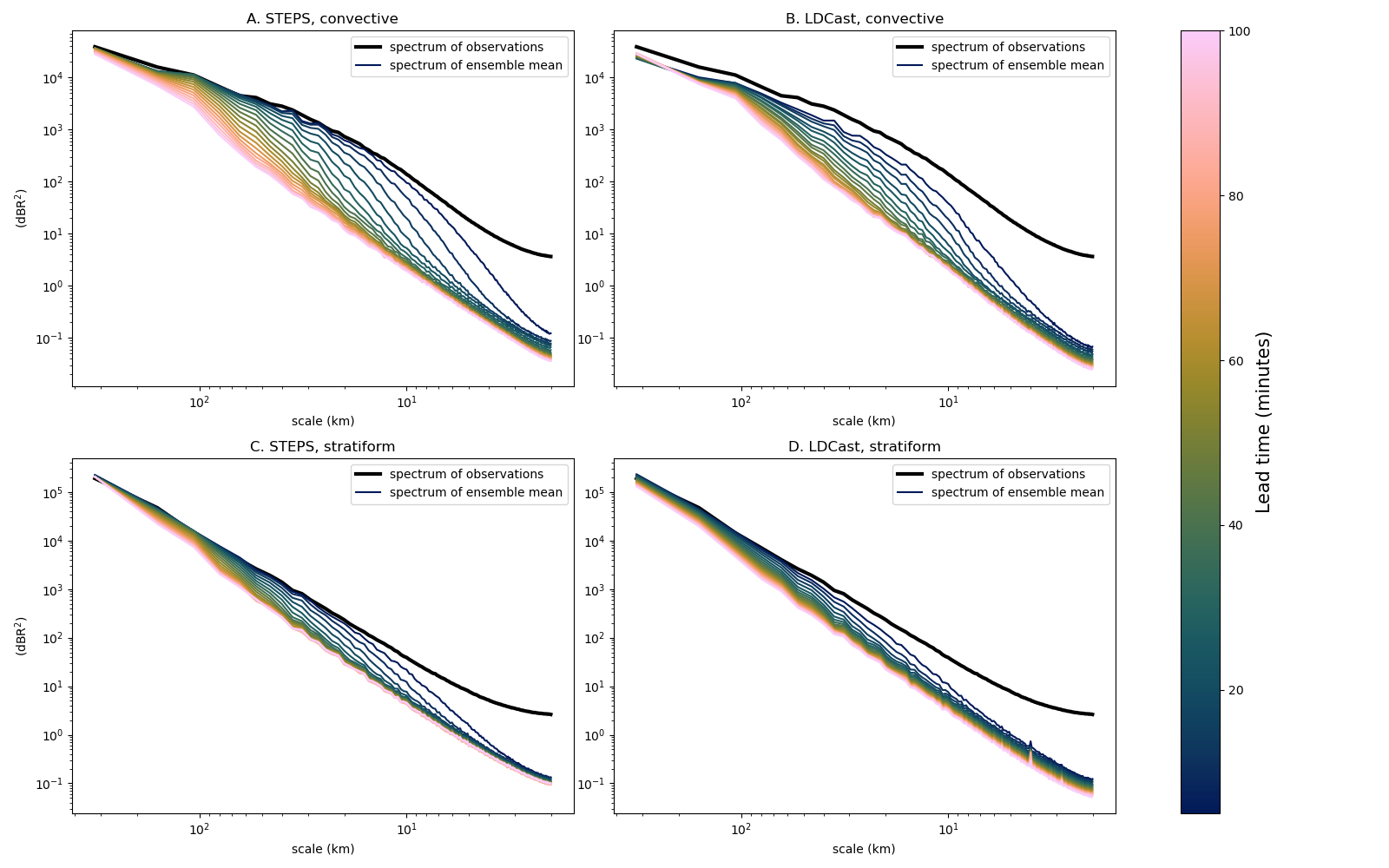}
\caption{Comparison of power spectra of ensemble means averaged over events (thin colored lines) with the mean power spectrum of observations (thick black lines) for convective and stratiform cases, and for STEPS and LDCast. The color represents the lead time.}
\label{spectrum_mean_fig}
\end{figure}

The spectral error is close to the power spectrum of observations because the ensemble mean has much less power at saturation than the observations (often one order of magnitude smaller, cf. Fig. \ref{spectrum_mean_fig}). This means that the Fourier coefficients of the ensemble mean are small with respect to those of the observation, so that the latter provide the main contribution to the error: $e\approx y$ (cf. Eq. \eqref{error_eq}).

For the spectral variance, at saturation, the power spectrum of the ensemble mean is small with respect to the power spectra of ensemble members, which themselves remain similar to that of the observation. All the terms in the sum in Eq. \eqref{spectral_variance_eq} are therefore also of the same order as the power spectrum of the observation.

Note that, when the error is computed as the difference between any two ensemble members, it is on average equal to twice the variance:
\begin{align}
\frac{1}{N(N-1)}\sum_{i, j = 1}^N PS(x_i - x_j)_k =& \frac{1}{N(N-1)}\sum_{i, j = 1}^N \langle|(X_{i, k} - \bar{X}_k) - (X_{j, k} - \bar{X}_k)|^2\rangle_{|\vec k| = k}\\
=& \frac{2}{N-1}\sum_{i = 1}^N \langle|X_{i, k} - \bar{X}_k|^2\rangle_{|\vec k| = k} - \frac{2}{N(N-1)}\sum_{i, j = 1}^N \langle(X_{i, k} - \bar{X}_k)^* (X_{j, k} - \bar{X}_k)\rangle_{|\vec k| = k}\\
=& 2 \sigma^2_k
\end{align}
The second term of the second line vanishes because $\sum_{i=1}^N(X_{i,k} - \bar{X}_k) = 0$. Therefore, if the error is taken to be the difference between two members (and not the difference between a member and the mean), it saturates in average to $2PS(x)_k$ (since $\sigma_k^2$ saturates to $PS(x)_k$).

\end{appendices}

\bibliographystyle{apalike}
\bibliography{biblio}

\end{document}